# THz Acoustic Attenuations of Glycerol by Pump Probe Technique

Fan Jun Wei[1], and Kung-Hsuan Lin[2]

[1] *Taiwan Instrument Research Institute, National Applied Research Laboratories, Hsinchu 30261, Taiwan*

[2] *Institute of Physics, Academia Sinica, Taipei 11529, Taiwan*

fanjw3@gmail.com, kunghsuan@gmail.com

The coherent acoustic waves can be created by InGaN double quantum wells. We measured the signals information of the coherent acoustic waves at the interface of GaN/glycerol to obtain the attenuations of glycerol at 295 K and 120 K by the femto-sec laser pump probe technique in THz regime. The attenuations of glycerol are calculated by using the acoustic mismatch model (AMM) through the complex acoustic impedance. We compared pump probe results with inelastic neutron scattering (INS) and inelastic X ray scattering (IXS), found they are matched within 0.5 THz at room temperature. The glycerol attenuations are found increasing monotonically to $1.75 \times 10^{8}\ m^{-1}$ and $1.3 \times 10^{8}\ m^{-1}$ till 0.5 THz by the pump probe technique at 295 K and 120 K, respectively.

## 1 INTRODUCTION

There are few methods to measure high frequency acoustics in the THz regime, such as inelastic X ray scattering[1] (IXS) or inelastic neutron scattering[2] (INS). But these two methods have their limitations, IXS has a strong quasi elastic scattering of the acoustic waves near 0 Hz, so the frequency could not be measured accurately less than 0.2 THz. INS need to bring a lot of theoretical models.[3,4] These models are pure theories and have not been verified in detail by experiments. Our method is very feasible in the THz regime, because the AMM could be utilized to study the acoustic attenuation in the THz regime and has been proved right in our previous paper.[5,6] The complex acoustic impedance, which includes the information of the dispersion relations and acoustic attenuations, is a parameter in the AMM to describe the phenomena when acoustic waves encounter the interface of media.[7] The mismatch of acoustic impedances results in reflection and transmission of acoustic waves at the interfaces. However, the sample must be thin enough that the intensity of transmission waves can be still measurable after they travel through the entirely coated film. But, sometimes, it is a challenge to prepare nano-meter-thick layers of a material for measurements such as liquid. In theory, the acoustic properties of a medium could be experimentally studied by the reflection of the acoustic waves.

The method of the reflection waves can open the field of studying the THz acoustic properties of liquid or glassy complex oxide. Because glycerol is liquid at room temperature, and the thickness of glycerol cannot be known by TEM or other methods. The solution of measuring a liquid or a glass is to measure the reflective property of the acoustic waves. However, there are still some interface scatterings between GaN and glycerol interface. We did twice process of the experiments of one sample to avoid interface effects. The first experiment of the sample was not coated with glycerol, and the second experiment was coated with glycerol, we used the laser light to hit the same spot on this sample, so that the GaN/glycerol interface effects could be eliminated.

## 2 EXPERIMENTS SETUP

By using the pump probe technique, we generated and detected the picosecond acoustic waves in the InGaN double quantum wells. In this work, we measured electronic responses of the interface information by using double InGaN double quantum wells as THz acoustic transducers. We do twice the experiments for a sample, this method can eliminate the interface effects of GaN/glycerol interface. Typical degenerate and noncollinear pump probe measurements were conducted with 390 nm optical pulses, which were frequency-doubled pulses from a Ti:sapphire oscillator. The polarization of the pump beam was orthogonal to that of the probe beam. The repetition rate of the pulses was 80 MHz. The diameter of the optical spots on the sample was measured to be approximately 28 $\mu$m. The pump and probe fluences were approximately 142 and 36 $\mu J/cm^2$, respectively. A polarizer was placed in front of the photodetector to eliminate pump light leakage. The pump beam was modulated at 2 MHz using an acoustic-optical modulator. A lock-in amplifier was used to record the transmission variation of the probe pulse as a function of time delay between the pump and probe pulses. With photoexcitation of femtosecond pulses, picosecond acoustic waves were generated in the InGaN double quantum wells. The transmission of the probe light was modulated when the acoustic waves travel through the double quantum wells.

Fig. 1 shows the schematic to indicate propagation and detection of generated acoustic waves from double quantum wells. After acoustic waves are generated in the quantum wells separately, they propagate in counter directions. The first peak feature around 7 ps corresponds to the generated acoustic waves leaving the quantum wells, and the third peak around 18 ps corresponds to the echoed acoustic waves in Fig. 2. The signals can be detected by the electronic changes in the double quantum wells when acoustic waves toward to the surface of the sample are partially reflected from the GaN/glycerol interface and go back to the quantum wells.

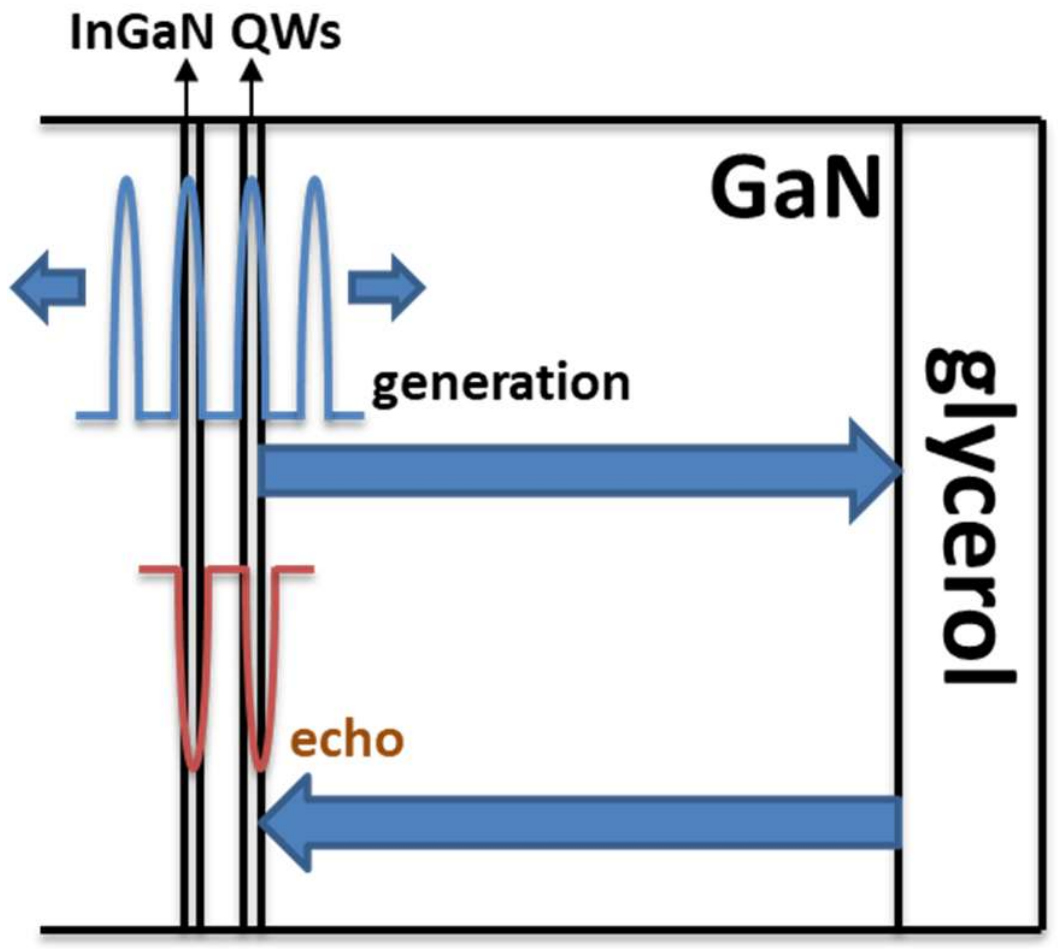

Fig. 1 The schematic to indicate propagation and detection of the acoustic waves of the generation and of the echo from double quantum wells.

# 3 DATA ANALYZTION

## 3.1 Raw Data

The raw data shows the transmission changes of the probe pulses as a function of time delay between pump and probe pulses. There are twice process of the experiments of one sample. The first experiment is the sample was not coated with glycerol. The raw data of the first experiment is shown in the black line of Fig. 2. Then we apply glycerol with a cover glass on the GaN cap layer of the same sample. The raw data of the second experiment is measured with the sample of the coated glycerol. The red line of Fig. 2 shows the raw data of the second experiment that the sample is coated with glycerol. On the background signals, which is due to the photoexcited carriers in the InGaN double quantum wells. We used the laser light to hit the same spot on the sample so that the GaN/glycerol interface effects could be eliminated by dividing the spectra in frequency domain. The theorical details will be described in the section 3.4.

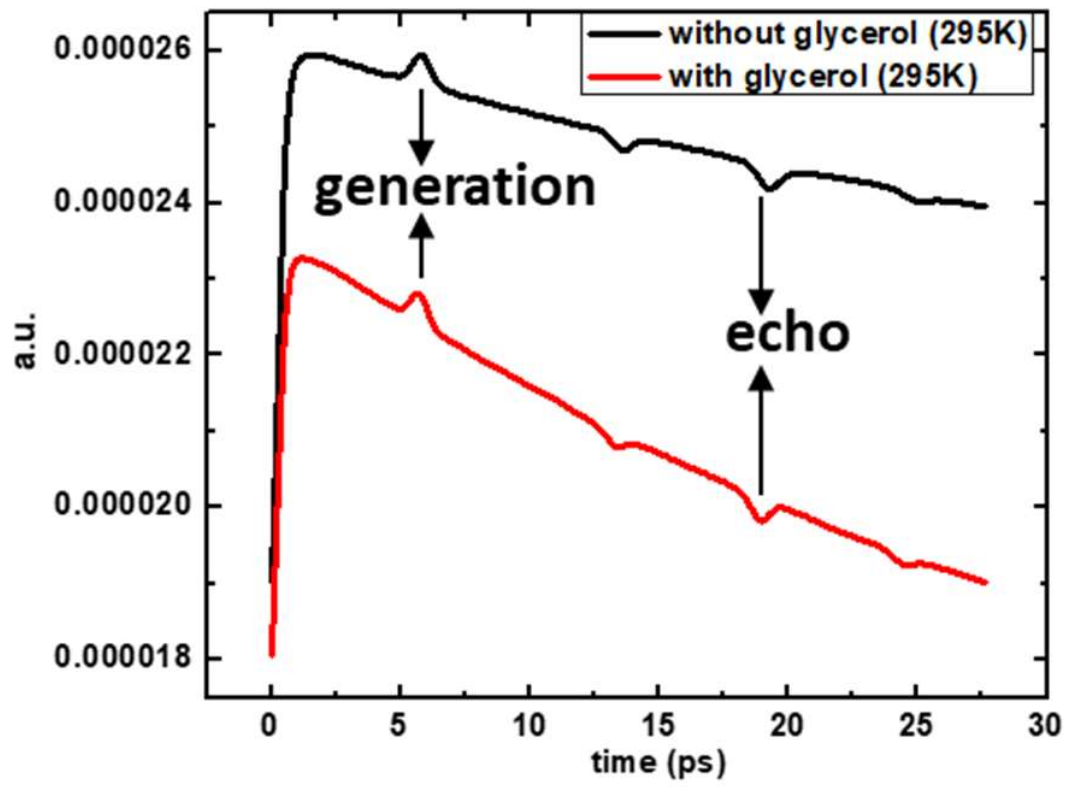

Fig. 2 The signals of the probe pulses as a function of time delay between pump and probe pulses. The black line is that the cap layer has not been coated with glycerol, the red line is coated with glycerol.

## 3.2 The Time Domain

The generation signals that are the background subtracted data of Fig. 2 are illustrated in Fig. 3, the black line is that the cap layer has not been coated with glycerol, and the red line is coated with glycerol. Similar with Fig. 3, Fig. 4 shows the echo signals that are the background subtracted data of Fig. 2. The linear functions were used to fit the background signals for subtraction. The time axes of the acoustic signals were individually shifted to zero time point for Fig. 3 and Fig. 4, and the sign of echoed signals from GaN/vacuum and from GaN/glycerol was reversed for comparison.

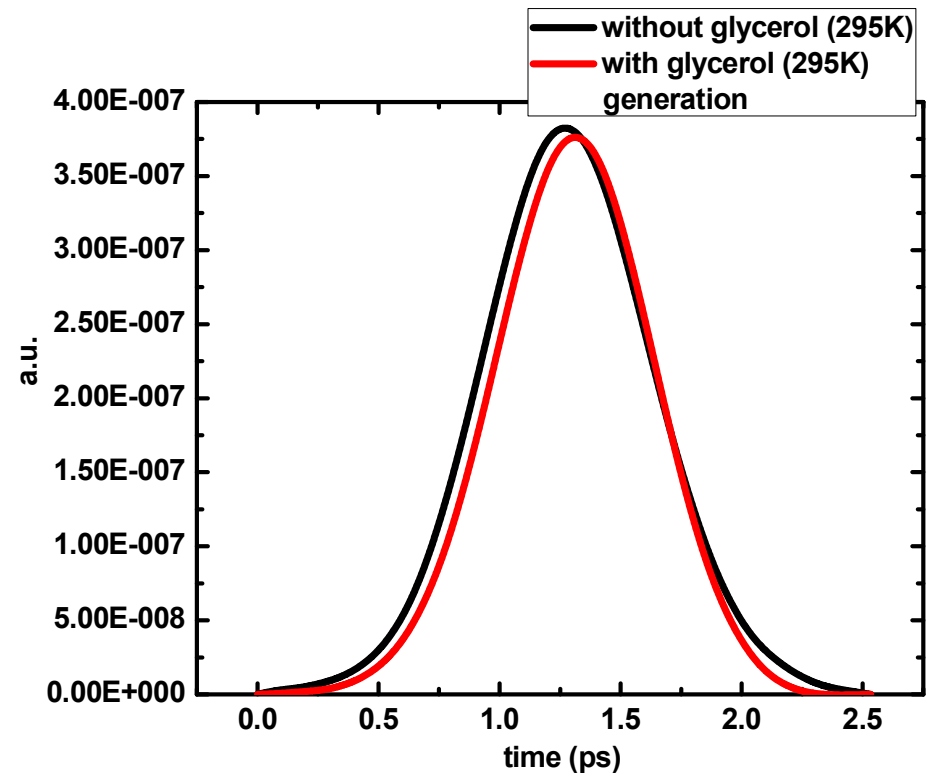

Fig. 3 The generation signals that are subtracted from the background of Fig. 2. The black line is that the cap layer has not been coated with glycerol, the red line is coated with glycerol.

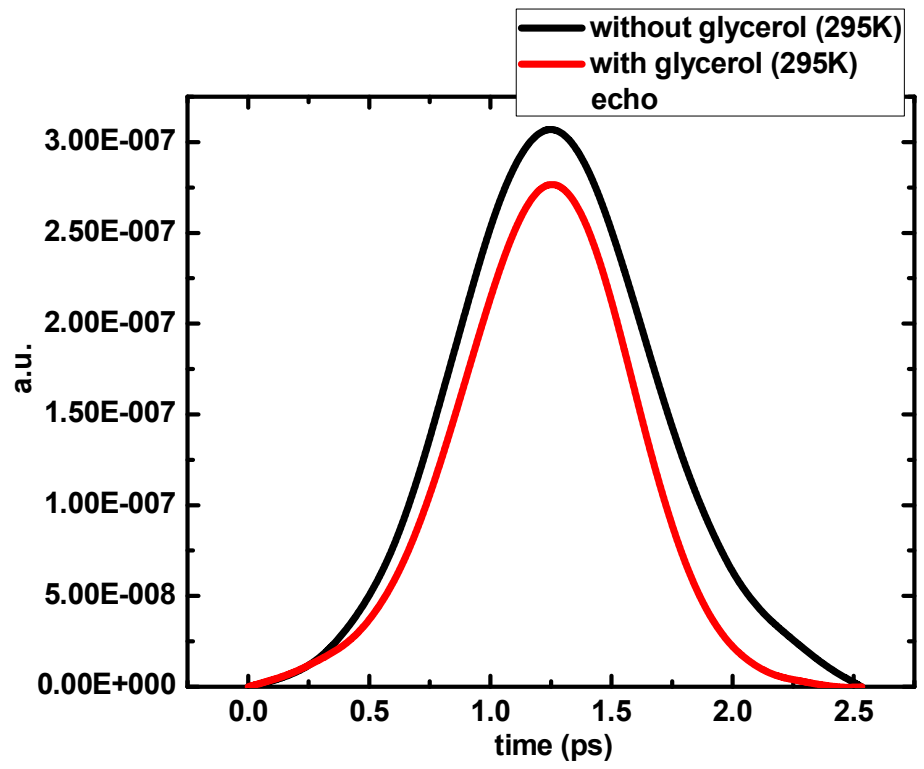

Fig. 4 The echo signals that are subtracted from the background of Fig. 2. The black line is that the cap layer has not been coated with glycerol, the red line is coated with glycerol.

## 3.3 The Frequency Domain

For Fig. 3 and Fig. 4, their corresponding spectra through Fourier transform are shown in Fig. 5 and Fig. 6. The black line is that the cap layer has not been coated with glycerol, the red line is coated with glycerol. Note that the four curves in Fig. 5 and Fig. 6 are the values used for AMM.

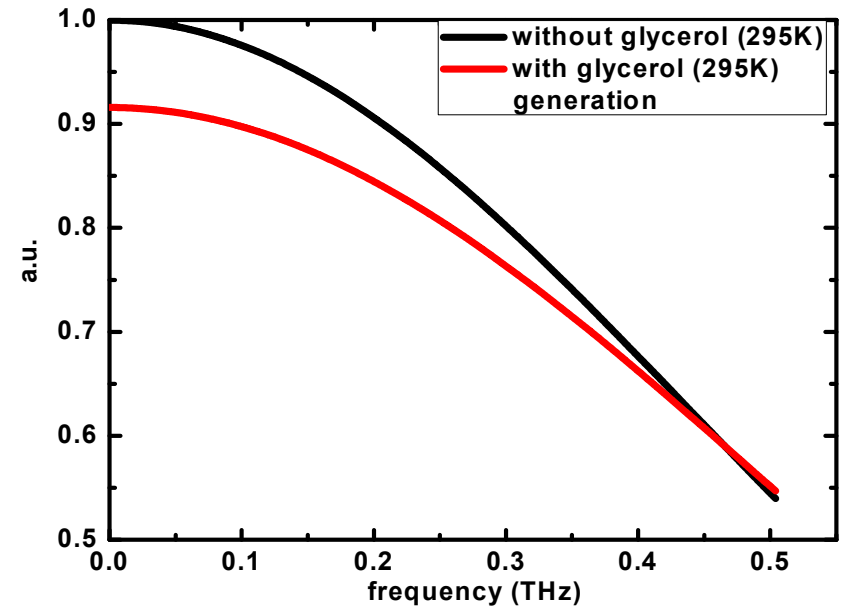

Fig. 5 The corresponding spectra in frequency domain of Fig. 3.

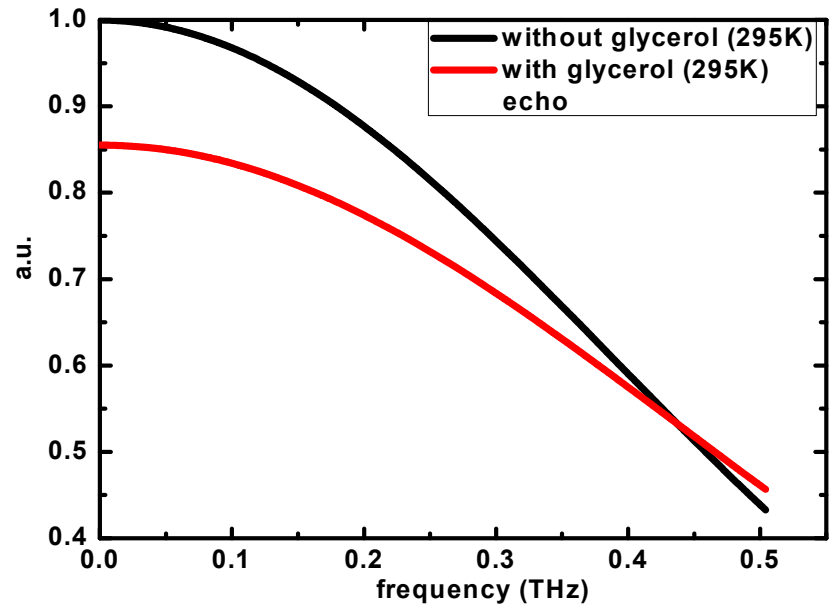

Fig. 6 The corresponding spectra in frequency domain of Fig. 4.

## 3.4 Acoustic Mismatch Model (AMM)

AMM is a theory about the acoustic waves propagate from one material to another material. In this paper, the acoustic waves propagate from the GaN to glycerol at room temperature and 120 K. The transmission and the reflection waves are produced at the interface. The acoustic impedance is a physical quantity to describe the traveling waves through AMM. The GaN/glycerol interface effects could be eliminated by dividing the spectra in frequency domain. The Eq. (1) is the complex form of AMM, the $\frac{E_g}{E_G}$ term is the reflectivity of GaN/glycerol interface, the $\frac{G_G}{G_g}$ term is used for calibrating the power of $\frac{E_g}{E_G}$ term, and the attenuations of glycerol is described in Eq. (2).[6]

$$r_r + ir_i = R = \frac{E_g G_G}{E_G G_g} \quad (1)$$

$$\alpha_g = \left(\frac{\alpha_G - \alpha_G r_r^2 - 2k_G r_i - \alpha_G r_i^2}{1 + 2r_r + r_r^2 + r_i^2}\right)\left(\frac{\rho_g}{\rho_G}\right) \quad (2)$$

Where $\alpha_g$ and $\alpha_G$ are attenuations of glycerol and GaN, respectively. $k_G$ is wave number of GaN, $\rho_g$ and $\rho_G$ are mass density of glycerol and GaN, respectively. $R$ is reflectivity. $r_r$ and $r_i$ are the real part and imaginary part of reflectivity. $G_g$ and $G_G$ are the generation of glycerol and GaN in the frequency domain in Fig. 5, respectively. $E_g$ and $E_G$ are the echo of glycerol and GaN in the frequency domain in Fig. 6, respectively.

## 3.5 The Results

Fig. 7. shows the glycerol attenuations raise monotonically to $1.75 \times 10^8\, m^{-1}$ and $1.3 \times 10^8\, m^{-1}$ within 0.5 THz by the pump probe technique at 295 K and 120 K, respectively. It can be notice that the attenuations increase rapidly after 0.2 THz, and the attenuation at 295 K is higher than at 120 K, this is in line with physical intuition that the lower temperature has the lower scattering rate corresponding to the attenuations.

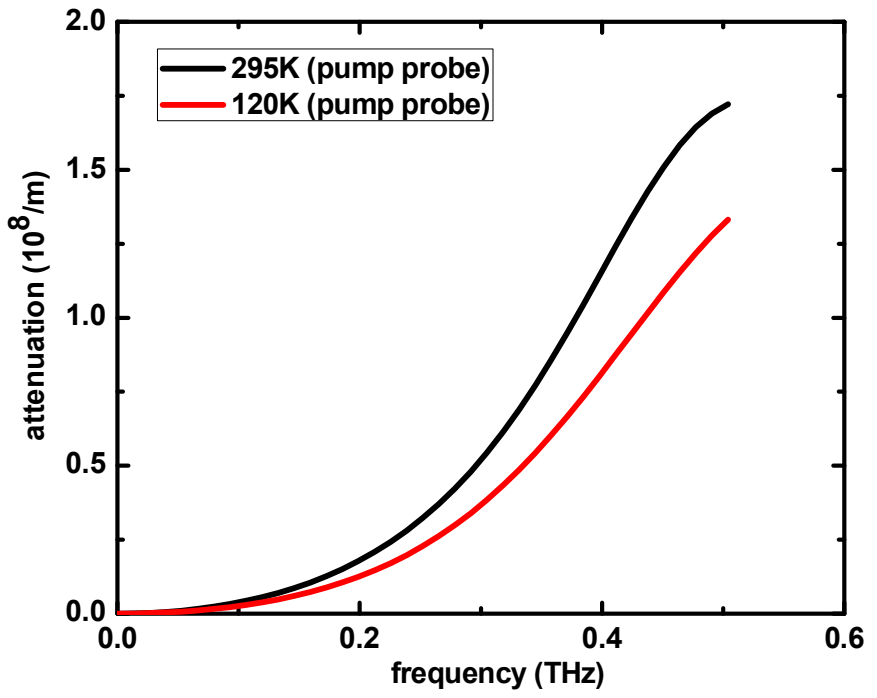

Fig. 7. The glycerol attenuations increase monotonically to $1.75 \times 10^8\, m^{-1}$ and $1.3 \times 10^8\, m^{-1}$ within 0.5 THz by the pump probe technique at 295 K and 120 K, respectively.

We compared the attenuations of different techniques by using AMM, IXS, and INS methods shown in Fig. 8. For AMM analysis, the acoustic waves do not really propagate through the entire glycerol film. The attenuation of glycerol is obtained by measuring the reflected acoustic waves at the GaN/glycerol interface. The glycerol attenuations are determined in the range of 0 to 0.5 THz by the pump probe technique plotted in the black line of Fig. 8. We found that our results at room temperature are matched the results of IXS and INS shown in Fig. 8.

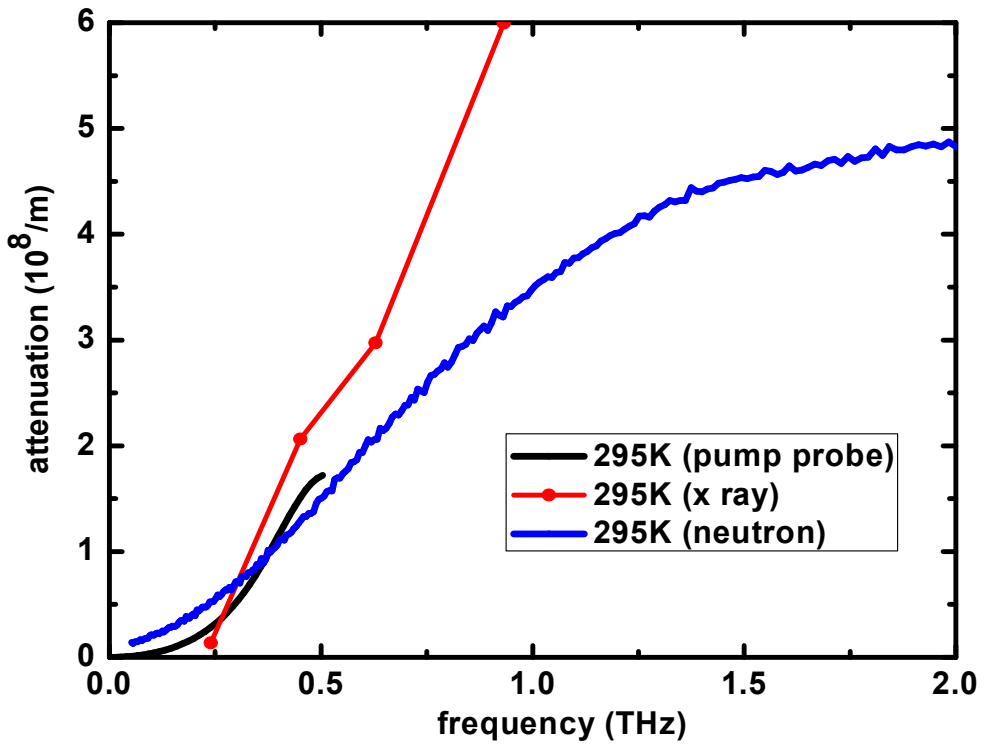

Fig. 8 The attenuations of the acoustic waves against the frequency in THz regime. The black line shows the pump probe method, the red line shows the results of IXS, and the blue line shows the results of INS method through the soft potential model.

## 4 CONCLUSIONS

In this work, we analyse the generated acoustic waves and the echoed acoustic waves between GaN and glycerol interface. The attenuations of glycerol at room temperature and 120 K based on AMM within 0.5 THz are obtained. The associated spectra in the frequency domain used for calculating the attenuations of glycerol are also measured. Overall, the results of attenuations by AMM, INS, and IXS are matched within 0.5 THz at room temperature. The glycerol attenuations are found increasing monotonically to $1.75 \times 10^8\ m^{-1}$ and $1.3 \times 10^8\ m^{-1}$ within 0.5 THz by the pump probe technique at 295 K and 120 K, respectively.